\documentclass{eas}
\usepackage{graphicx}
\begin{document}

\title{On the Prompt Signals of Gamma Ray Bursts} 
\author{Pisin Chen}
\address{Department of Physics and Leung Center for Cosmology and Particle Astrophysics (LeCosPA), National Taiwan University, Taipei, Taiwan 10617 \& KIPAC, SLAC, Stanford University, CA 94035, U.S.A.  \email{pisinchen@phys.ntu.edu.tw}}
\author{Toshi Tajima}
\address{ZEST \& Ludwig-Maximilians-Universitat Munchen, Fakultat f. Physik, am Coulombwall 1, D-85748 Garching, Germany}
\author{Yoshi Takahashi}
\address{Posthumous; Department of Physics, University of Alabama, Huntsville, AL 35899, U.S.A.}

\runningtitle{GRB prompt signals}
\begin{abstract}
We introduce a new model of gamma ray burst (GRB) that explains its observed
prompt signals, namely, its primary quasi-thermal spectrum and high energy
tail. This mechanism can be applied to either assumption of GRB progenitor:
coalescence of compact objects or hypernova explosion. The key
ingredients of our model are: (1) The initial stage of a GRB is in the form
of a relativistic quark-gluon plasma Òlava"; (2) The expansion and cooling of
this lava results in a QCD phase transition that induces a sudden gravitational
stoppage of the condensed non-relativistic baryons and form a hadrosphere;
(3) Acoustic shocks and Alfven waves (magnetoquakes) that erupt in
episodes from the epicenter efficiently transport the thermal energy to the
hadrospheric surface and induce a rapid detachment of leptons and photons
from the hadrons; (4) The detached $e^+e^-$ and $\gamma$ form an opaque, relativistically
hot leptosphere, which expands and cools to $T \sim mc^2$, or 0.5 MeV,
where $e^+e^- \to 2\gamma$ and its reverse process becomes unbalanced, and the GRB
photons are finally released; (5) The Òmode-conversion" of Alfven waves into
electromagnetic waves in the leptosphere provides a ÒsnowplowÓ acceleration
and deceleration that gives rise to both the high energy spectrum of GRB and the erosion of its thermal spectrum down to a quasi-thermal distribution. According to this
model, the observed GRB photons should have a redshifted peak frequency
at $E_p \sim \Gamma(1 + \beta/2)mc^2/(1 + z),$ where $\Gamma\sim {\cal{O}}(1)$ is the Lorentz factor of the
bulk flow of the lava, which may be determined from the existing GRB data.
\end{abstract}
\footnotetext{Originally released as SLAC-PUB-8874 in 2001, this paper was never formally published.
}
\maketitle
\section{Introduction}

The Òfireball modelÓ of GRB proposed in 1980s (Paczynski \cite{Paczynski1986}; Goodman \cite{Goodman1986}; Shemi \& Piran \cite{Shemi1990}), 
which assumes a smooth
expansion of the fireball, was later regarded as having difficulty to produce the high
energy tail of the spectrum (Rees and Meszaros \cite{Rees-Meszaros}, Meszaros and Rees \cite{Meszaros-Rees}). This difficulty arises from the issue of Òbaryon loading,Ó
where light particles (such as photons and electrons/positrons as well as neutrinos)
cannot be easily detached from the opaque baryonic matter. It is generally believed
that such a system would convert most of its energy into kinetic energy of the baryons
rather than the luminosity. Indeed, Rees and Meszaros (\cite{Rees-Meszaros},\cite{Meszaros-Rees}) focused on this feature as
the major issue of GRB and proposed an alternative Òfireball shock model.Ó In this
model the exploding $e^+e^-$ plasma has a bulk Lorentz factor $\Gamma \sim 10^2-10^3$ at a radius
of $\sim 10^5$ km. While this model addresses the issue of high energy tails, with a large
Lorentz factor it remains a challenge to explain the spectral peak at several hundred keV.

The typical spectrum of a GRB consists of a relatively broad, thermal-like spectrum,
with the peak energy $E_p$ located at a few hundred keV, which contributes more than
half of its total luminosity. In the illustrative case of GRB 990510, activities of low energy
spectrum ($< 62$ keV) precede the main sudden onset of the high energy spectrum
($> 330$ keV) by a few 10s of seconds. In addition to the spectrum around the peak,
a substantial fraction of the total luminosity is contributed from the high energy tail,
which can be characterized by a power-law with a (negative) index $\sim 2-2.5$. In terms of
the time structure, GRBs can be classified into two types: the short bursts that last for
$\sim$ 1-10 sec and the long bursts that last for tens to hundreds of seconds. It is interesting
to note that while the time duration and profile vary widely over several orders of
magnitude, the GRB spectra described above are remarkably universal. Much attention
has been devoted to analyzing GRB afterglow as a result of an expanding fireball, which
leads to important correspondence between the observational data and phenomenological
models. However, a comprehensive understanding of the underlying mechanisms
that produces such a fireball and the prompt signals are still lacking.

We suggest that the key to the understanding of GRB lies in its prompt signals,
in particular the thermal portion of the spectrum. In this article we propose a new
GRB model which provides a unified picture on the early-stage evolution and thus the
mechanism that produces the prompt signals of GRBs. The key ingredients of our model
are:

1. In the final stage of either compact-object coalescence or hypernova explosion,
large fragments of hadron matter are ejected, most likely non-isotropic. Heated by the
release of a large fraction of the systemÕs gravitational potential energy, the hadrons are
melted into quarks and gluons with temperature $\sim 200$MeV and density $10^{38}{\rm cm}^3$,
like a molten lava. The bulk flow of such a Òlava,Ó or hadrosphere, however, is only
mildly relativistic.

2. The expansion and cooling of this lava results in a QCD (quantum chromodynamics)
phase transition at a temperature $\sim 120$MeV and density $\sim 2 \times 10^{37} {\rm cm}^3$ that
condensates the relativistic quarks and gluons into non-relativistic baryons. These nonrelativistic
baryons feel the strong gravity and stop their expansion. This results in the
formation of a ÒhardenedÓ hadrosphere boundary, analogous to the darkening of the
lava surface.

3. Acoustic shocks and Alfven waves (magnetoquakes) that erupt in episodes from
the epicenter efficiently transport the thermal energy to the hadrospheric surface and
induce a rapid detachment of leptons and photons from the hadrons.

4. The detached $e^+e^-$ and $\gamma$ form an opaque, relativistically hot leptosphere, which
expands and cools to $T \sim mc^2$, or 0.5 MeV, below which $e^+e^-\to 2\gamma$ and its reverse
process become unbalanced, and the GRB thermal photons are released. The observed
peak of this portion of the GRB spectrum is $E_p \sim \Gamma (1+\beta/2)mc^2/(1+z)$, where $\Gamma \sim {\cal{O}}(1)$
is the Lorentz factor for the bulk flow of the Òlava,Ó and $z$ is the GRB redshift factor.

5. The existence of a nonlinear $e^+e^-$ plasma-mediated Òmode-conversionÓ effect
that converts Alfven waves into electromagnetic waves in the leptosphere. This process
provides a novel ÒsnowplowÓ acceleration and deceleration mechanism that produces both the high energy
spectrum of GRB and the erosion of its thermal spectrum down to a quasi-thermal distribution.

Figure 1 is a schematic diagram that depicts our GRB model. In the following sections we elaborate these key points of our model in more details.

\begin{figure}[htb]
  \begin{center}
    \includegraphics[height=5cm]{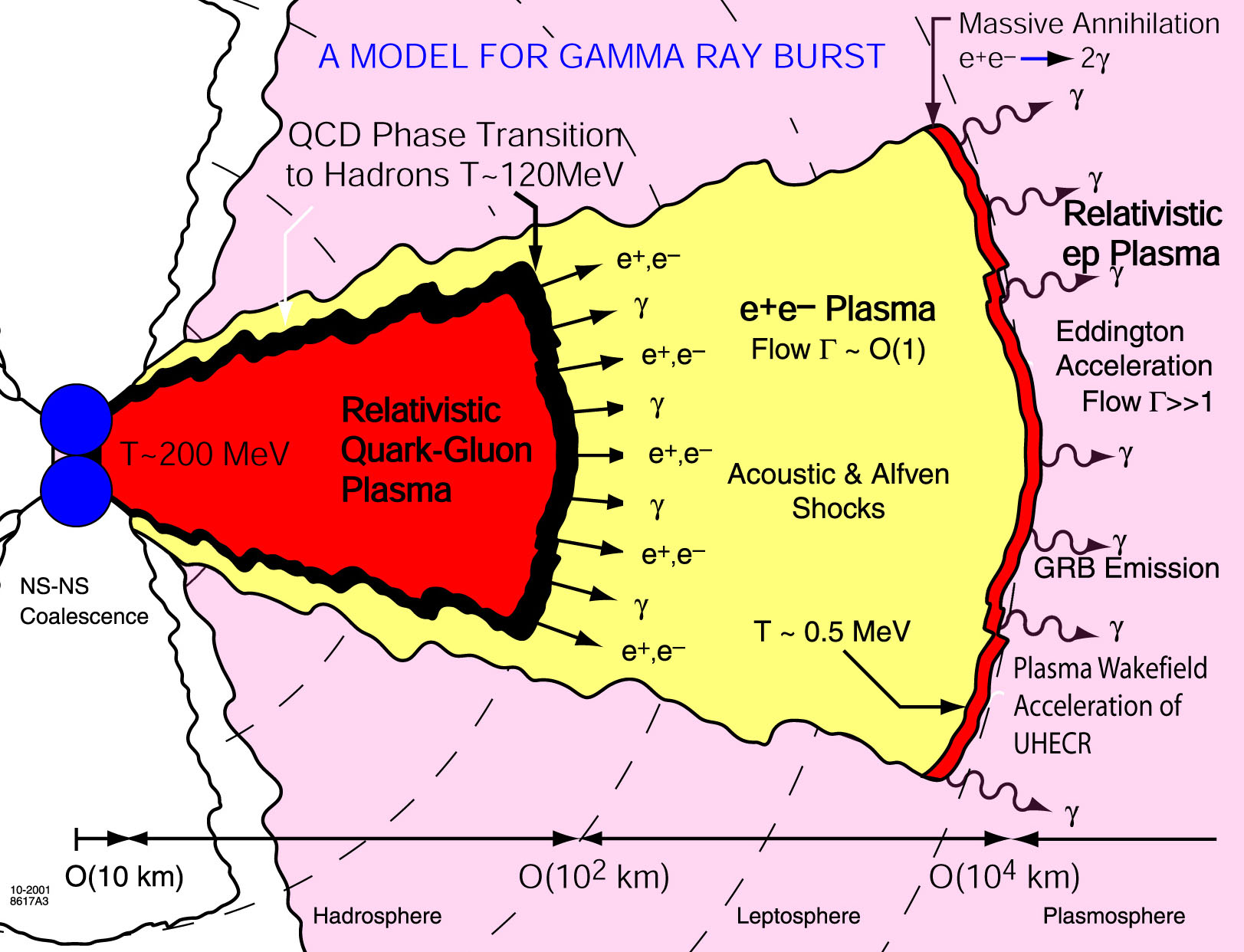}
  \end{center}
 \vspace{-1.pc}
  \caption{A schematic diagram that depicts the various phases of the GRB dynamics in our model in the aftermath of the coalescence of a binary neutron star system.}
\label{fig:1}
\end{figure}

\section{Hadrosphere and QCD Phase Transition}

We assume that in the final stage of either compact-object coalescence or hypernova
explosion, the tremendous concentration of energy triggers the eruption of large fragments
of baryon matter. The density of baryon matter under such circumstance is
comparable to that of a neutron star, i.e., $\sim 10^{38} {\rm cm}^3$. Heated by the system's released
gravitational energy, which can be as large as $\sim 0.1-0.3$ of the total rest mass of the
system, such baryon fragments can gain a thermal energy, or temperature, $\sim 200$MeV.
Under high temperature and density, one expects from quantum chromodynamics
(QCD) that the baryon matter turns into a deconfined quark-gluon plasma (Alford \cite{Alford1998}). A quantitative
description of such QCD phase transition has been a major challenge to nuclear
physicists. The standard approach is to invoke grand canonical
ensemble (in which the particle number is not fixed), and therefore the relation between
the temperature and the chemical potential. Nevertheless, we believe that the
phase transition happens at temperature $T \ge 120$MeV for ÒzeroÓ baryon and at density
$\rho \ge 10^{39} {\rm cm}^3$ for much lower temperature (Liu \cite{Liu2001}). Taking these conditions as our constraint
and translating the chemical potential into an average particle density, we can
parameterize the QCD phase boundary as
\begin{equation}
\Big(\frac{\rho}{\rho_c}\Big)^2 + \Big(\frac{T}{T_c}\Big)^2 =1,
\end{equation}
where $\rho_c \sim 10^{39}{\rm cm}^3$ and $T_c \sim 120$MeV. Clearly, the initial state of our system is in the
quark-gluon phase.

Once quarks are deconfined at such energy-density, they are highly relativistic since
their rest masses are as low as $m_u \sim 4$ MeV and $m_d \sim 7$ MeV for the up and down
quarks, respectively. In this plasma there are about the same order of magnitude in the
electron/positron (and neutrino) populations as well as thermal energies, since
they are in (near) local thermal equilibrium with the relativistic quarks and gluons.
Once this ÒlavaÓ of quark-gluon plasma erupts, it adiabatically expands and cools. We
call such a cluster the hadrosphere.

As the hadrosphere expands to the radius of $\sim 50$ km, the quark-gluon plasma
density reduces to $\rho_{q-g} \sim 2 \times 10^{37}{\rm cm}^3$. From thermodynamics the temperature and
density are related by
\begin{equation}
\rho^{1-\gamma}T= {\rm const.}
\end{equation}
For relativistic particles, $\gamma = 4/3$, and we find $T \propto \rho^{1/3}$. Since $\rho\propto 1/V \propto 1/R^3$, we
have $T \propto 1/R$. Thus the temperature drops to $T \sim 120$MeV at this point. This is the temperature
for QCD phase transition when the density is much lower than the critical one: $\rho\ll
\rho_c$. Note, however, that in the case of NS-NS coalescence, the initial baryon density would be much higher than that of the nucleus, and therefore a much larger chemical potential. $T_c$, which is a function of both temperature and chemical potential, is thus much lower, at $\sim 10-20$MeV, and therefore the QCD phase transition is easier to reach. Once $T \sim T_c$, the quarks and gluons condensate into hadrons and turn nonrelativistic, which immediately feel
the immense gravity and are thus gravitationally trapped. This gravitational capture of
baryonic matter marks the boundary of the hadrosphere.

Note that such a quark-gluon explosion needs not be spherically symmetric, and may
be irregular or even in chunks. Under the extreme high densities, the hadrosphere is
highly opaque and poor in convection. Thus the quark-gluon plasma near the boundary
first condensate into baryons while its interior is still molten. This is analogous to the
darkening of the lava surface after erupted from the volcano, where the interior of the
lava is still red-hot.

\section{Separation of Photons and Leptons from Hadrosphere}

As mentioned in the Introduction, one seeming difficulty in the fireball model is the lack
of a mechanism to efficiently transport the tremendous luminous energy near instantly
across the baryonic matter. Given the extremely high density and therefore short mean-free-path 
in the fireball, the transport of energy through individual particle kinematics,
i.e., thermal convection, would indeed be hard. This is the well-known problem of
Òbaryon loading.Ó It may be overcome, however, by the transport of energy through
collective plasma excitations.

In the final stage of compact-object coalescence or the collapse of supermassive
star we expect the generation of strong acoustic waves (Òinternal shocksÓ) and Alfven
waves (we may call this ÒmagnetoquakesÓ). These waves are efficient mass and energy
carriers (Holcomb \& Tajima \cite{Holcomb}) in the interior of the hadrosphere as well as the leptosphere. For example,
in the NS-NS or NS-BH coalescence, the violent perturbations of the strong magneticfield
pressure of the host neutron stars ($B \sim 10^{12}-10^{13}$G) induces the excitation of
magnetoquakes. As much as $\sim{\cal{O}}(10^{52})$ erg of energy may be carried by these waves.
Due to the compactness of the progenitor, the period of these magnetoquakes is about
$\sim {\cal{O}}(100)\mu$sec during each episode. As these shocks approach the boundary of the
hadrosphere, the tortional as well as the compressional Alfven waves in the rapidly
density-graded stellar magnetosphere are expected to exhibit interesting and important
properties (Takahashi, Hillman, and Tajima \cite{Takahashi2000}). One is precisely 
the possibility of transport of energies from the epicenter
to the hadrosphere boundary during each episode of magnetoquake. Another
is the possibility of Òmode-conversionÓ in the leptosphere. The density of the leptospheric
$e^+e^-$ plasma decreases rapidly due to its expansion. In such an environment
the torsional Alfven waves can mode-convert themselves into the usual electromagnetic
waves (Daniel and Tajima \cite{Daniel1998}).

At the surface of hadrosphere, where the non-relativistic baryons are suddenly slowed
down by self-gravitation as a result of QCD phase transition, the still highly relativistic
leptons are freely radiating through the surface and their chemical potentials are negligible.
Thus in close analogy with the standard blackbody emission process and according to the Stefan-Boltzmann law, $J = \alpha T^4$,
where $\alpha = 5.67\times 10^5 {\rm erg/sec/cm}^2/K^4,$ the system can emit above $10^{52}$ ergs in $10^{-8}$ second
with a temperature of 120 MeV and a radius of 50 km.

\section{Mode-Conversion in Leptosphere}

As mentioned earlier, the emitted $e^+e^-$ and $\gamma$ are so dense that they are not freely
propagating outward. With tremendous near-instant supply of electrons and positrons,
the radiated $e^+e^-$ pairs (as well as photons and neutrinos) will likely create shocks.
As mentioned in the previous section, the internal acoustic and Alfvenic shocks can
provide efficient energy transport as well as snowplow acceleration within the dense
hadrosphere. In addition, the Alfven waves that continue to propagate across the leptosphere
can induce a novel, linear and nonlinear phenomenon called Òmode-conversion.Ó
The density of the leptospheric $e^+e^-$ plasma decreases rapidly due to its expansion. It
has been observed in the particle-in-cell computer simulations that in such an environment
the torsional Alfven waves can mode-convert themselves into ordinary electromagnetic
waves (Kippen \cite{Kippen1999}). Furthermore, it was observed that inside such an opaque plasma
a Òself-induced transparencyÓ occurs. Namely, a large number of energetic particles
are plowed and accelerated in front of the Alfven wave, which are detached from the
opaque, collisional bulk plasma. 

When the mode-conversion occurs in the $e^+e^-$ plasma, the converted EM waves proceed
ahead of the Alfven waves and the snowplowed particles, forming an integrated
overall trinity structure. This structure is capable of converting a large fraction of the
wave energy (magnetoquake energy) into kinetic energies of the accelerated particles,
as well as the heating of the bulk plasma. In our scenario this mechanism provides the
basis of the production of the nonthermal high energy spectrum of GRB. The mechanism
of this transport is analogous to ÒsnowplowingÓ: particles are pushed forward in
front of the shock waves. We note that such process can also decelerate those particles that are on the `wrong side' of the slope between episodes of 
magnetoquakes (Chen, Tajima, \& Takahashi \cite{Chen2002}). Such stochastic processes will dilute the pure thermal spectrum into a quasi-thermal one.  
  
\section{Quasi-Thermal Spectrum of GRB}

By the time when the leptosphere expands to a radius $\sim$ 10,000 km and cooled to below
the two-photon pair production threshold, i.e., $T \sim mc^2 \sim 0.5$ MeV, the two-photon
pair production and its reversed pair annihilation processes,
\begin{equation}
e^+e^-\to 2\gamma\,
\end{equation}
are out of balance, and the $e^+e^-$ are largely annihilated into photons with a typical energy
of $E_{p0} \sim 0.5$ MeV in the rest frame of the bulk flow.
The observed peak energy of the GRB thermal spectrum should therefore be 
\begin{equation}
E_p \sim \frac{\Gamma(1+\beta/2)}{1+z}E_{p0}\sim \frac{\Gamma(1+\beta/2)}{1+z}mc^2,
\end{equation}
where $z$ is the GRB redshift factor, $\beta^2=1-1/\Gamma^2$ and $\Gamma$ is the Lorentz tactor of the bulk flow of the lava.
As explained above, such initially thermal spectrum will be eroded to a quasi-thermal one due to the stochastic nature of snowplow acceleration-deceleration interplay under random phases of magneto quakes or shock waves. 

To compare our model with observations, we take long burst GRBs with redshift factors identified from Piran, Jimenez, and Band (\cite{Piran2000}), based on the BATSE data. There are 8 events where both the redshift and the spectral peak, $E_p$, have been identified. Among these 8 events on, GRB980425, is discarded because it is very local ($z\sim 0.01$) and its total luminosity fell sufficiently below the typical GRBs.  

\begin{table}[h]
  \begin{center}
    \begin{tabular}{||c||c|c|c|c||}
      \hline
      ${\rm Burst Name}$ & $E_p^{obs} [{\rm keV}]$ & $z$ & $mc^2/(1+z)$ & ${\rm Derived}$ $\Gamma$  \\
      \hline
      ${\rm GRB970508}$ & $481$ &$0.84$ &$278$ & $1.3$ \\
      ${\rm GRB970825}$ & $230$ &$0.96$ &$261$ & $0.9$ \\
      ${\rm GRB971214}$ & $156$ &$3.41$ &$116$ & $1.1$ \\
      ${\rm GRB980703}$ & $370$ &$0.97$ &$259$ & $1.2$ \\
      ${\rm GRB990123}$ & $550$ &$1.60$ &$197$ & $1.9$ \\
      ${\rm GRB990506}$ & $450$ &$1.20$ &$232$ & $1.4$ \\
      ${\rm GRB990510}$ & $174$ &$1.62$ &$195$ & $0.9$ \\
      \hline
    \end{tabular}
    \caption{Comparison of our model with observations based on the 7 GRB events from the BATSE catalog.}
  \label{tabletime}
  \end{center}
\end{table}

\section{Conclusion}
 
We have discussed the key features of our new model for GRB. Our scenario appears
to be able to provide an explicit physical framework that can explain many of the GRB
quasi-thermal spectrum characteristics. These include the release of $\sim10^{52}$ erg of energy from
a compact source, the promptness of such a release, and the origin of the GRB spectral
peak as well as the high energy tail. Episodes of vibrations and eruptions of acoustic
shocks and magnetoquakes, which should have a period of $\sim 100\mu {\rm sec}$ during each burst,
induce a fine structure within the overall duration of the prompt GRB signals. We have not discussed 
the physics in the outer plasmosphere (which is formed beyond the boundary of the leptosphere 
where positrons are essentially all annihilated).
The existence of the plasmosphere, however, is in our view
essential to another very important astrophysical phenomenon, namely the production
of ultra-high energy cosmic rays (UHECR) beyond $10^{20}$ eV. 
 
\section{Acknowledgement}
This work was supported by US
Department of Energy, contract DE-AC03-76SF00515 (PC), DE-FG03-96ER40954 (
TT with UTA), W-7405-ENG-48 (TT with LLNL); and DE-FG-02-88ER41058 (YT);
and by NASA, contract NAS8Ð98226 (YT).



\begin{thebibliography}{99}
\bibitem[1986]{Paczynski1986}Paczynski, B., 1986, ApJ {\bf 308}, L43.
\bibitem[1986]{Goodman1986}Goodman, J., 1986, ApJ {\bf 308}, L47.
\bibitem[1990]{Shemi1990} Shemi, A., and Piran, T., 1990, ApJ {\bf 365}, L55.
\bibitem[1992]{Rees-Meszaros} Rees, M.J.,  and Meszaros, P., 1992, MNRAS {\bf 258}, 41. 
\bibitem[1993]{Meszaros-Rees} Meszaros, P., and Rees, M.J., 1993, ApJ {\bf 405}, 278.
\bibitem[1998]{Alford1998} Alford, M., ``New Possibilities for QCD at Finite Density", 1998, hep-lat/9809166.
\bibitem[2001]{Liu2001} Liu, Keh-Fei, private communications, April 2001.
\bibitem[1991]{Holcomb} Holcomb K.A., and Tajima, T., 1991, ApJ {\bf 378}, 682.
\bibitem[2000]{Takahashi2000} Takahashi, Y., Hillman, L.W., and Tajima, T., in {\it High-Field Science}, eds. Tajima, T.,
Mima, K., and Baldis, H., 2000, p. 171 (Kluwer Academic, NY).
\bibitem[1998]{Daniel1998} Daniel, J.,  and Tajima, T., 1998, ApJ {\bf 498}, 296.
\bibitem[1999]{Kippen1999} Kippen, R.M., 1999, GCN 322.
\bibitem[2002]{Chen2002} Chen, P., Tajima, T., Takahashi, Y., 2002, Phys. Rev. Lett. {\bf 89}, 161101.
\bibitem[2000]{Piran2000} Piran, T., Jimenez, R., and Band, D., ÒThe Energy Distribution of GRBs,Ó in Gamma-
Ray Bursts: 5th Huntsville Symposium, ed. R. M. Kippen, et al., 2000, AIP Porc. 1-56396-947.
\end{thebibliography}
\end{document}